\documentclass[amssymb,preprint,aps]{revtex4}

\usepackage{graphicx}

\begin{document}

\enlargethispage*{80pt}

\title{Statics and Dynamics of Condensed DNA
within \\ Phages and Globules }
\author{Theo Odijk} 
\email[E-mail: ]{odijktcf@wanadoo.nl}
\thanks{\newline
Address for correspondence: \newline
T. Odijk, P.O. Box 11036, 2301 EA Leiden, the Netherlands} 
\affiliation{Theory of Complex Fluids
\\ Faculty of Applied Sciences \\ Delft University of Technology \\
Delft, the Netherlands }

\begin{abstract}
Several controversial issues concerning the packing of linear DNA
in bacteriophages and globules are discussed. Exact relations for
the osmotic pressure, capsid pressure and loading force are
derived in terms of the hole size inside phages under the
assumption that the DNA globule has a uniform density. A new
electrostatic model is introduced for computing the osmotic
pressure of rodlike polyelectrolytes at very high concentrations.
At intermediate packing, a reptation model is considered for DNA
diffusing within a toroidal globule. Under tight packing
conditions, a model of Coulomb sliding friction is proposed. A
general discussion is given of our current understanding of the
statics and dynamics of confined  DNA in the context of to the
following experiments: characterization of the liquid crystalline
phases, X-ray scattering by phages, osmotic stress measurements,
cyclization within globules and single-molecule determination of
the loading forces.
\end{abstract}


\maketitle

\section*{1. Introduction}

The compaction of DNA in biological cells and viruses presents us
with rather unconventional problems in the physics of soft matter.
In the case of phages or viruses, the DNA may be so close-packed
that the thermodynamics is no longer extensive because it is
dominated by the energy arising from regions of tight bending
(Odijk 1998). Thus we are dealing with a defect rather than a bulk
system. Confronting the mathematical physics of the DNA
configurations as such is a formidable undertaking. Still, the
condensed DNA globule ought to have relatively minor fluctuations
in the density (Ubbink \& Odijk 1996) so that simplified analyses
are possible.

A general maxim in theoretical biology is not to theorize before
one has assimilated pertinent data and the same would seem to
apply to the field of DNA compaction also. In setting up the
computations outlined here, I have relied heavily on what we have
learned from a number of recent experiments. The spoollike
structure of DNA in T7 and T4 bacteriophages has been established
in a quantitative fashion (Cerritelli et al 1997; Olson et al
2001). For the phi29 phage we now have a fairly good comprehension
of the connector motor (Simpson et al 2000) and the typical forces
involved in DNA packaging as a function of length reeled in (Smith
et al 2001). Cyclization experiments (Jary \& Sikorav 1999) of DNA
in condensed globules allow us to estimate the local friction the
DNA coil experiences in a congested environment.

Provisional theories for DNA compacted in phages were already
proposed a couple of  decades ago (Reimer \& Bloomfield 1978;
Gabashvili et al 1991; Gabashvili \& Grosberg 1992). Nevertheless,
the notion that the tight curvature of the DNA sheath surrounding
a hole determines the interaxial spacing, was stressed only
recently (Odijk 1998). Theoretical activity in this area is
currently gaining momentum but will be discussed in the last
section. The experimental literature on DNA packaging in phages
and viruses is vast and complicated and, owing to space
limitations, cannot be dealt with here (for recent reviews of the
experimental situation, see Grimes et al (2002) and Jardine \&
Anderson (2003)). Here I wish to emphasize several theoretical
issues that have not been focussed on before.

A continuum approximation of the curvature energy of DNA condensed
within a capsid is derived in terms of the inhomogeneous DNA
density in the limit of small fluctuations. A further useful
approximation for close packed DNA is to let the density be
constant. It turns out that it is then possible to derive general
nanothermodynamic relations for quantities like the osmotic
pressure, force of insertion, etc. Owing to the very high degree
of packing in phages, we are obliged to reinvestigate the
classical cell model for rodlike polyelectrolytes. The
interpretation of the cyclization experiments of linear globular
DNA leads to the (surprising?) conclusion that simple reptation
seems to be valid though I note that regions of tight curvature
are often absent in globules. The friction of DNA in highly
congested phages is discussed in terms of current thinking on
sliding friction at nanoscales. In the last section, I confront
these new computations as well as previous theoretical work with
the experiments mentioned above.

\section*{2. Continuum approximations}

The persistence length of DNA is generally of the order of
magnitude of the typical size of bacteriophages. This makes the
statistical treatment of DNA viewed as a confined wormlike chain
with self-interaction very difficult. Fortunately, there is one
limit where we can make considerable headway, viz. when the DNA is
close-packed at moderate to high densities owing to bending
stresses. Such a condensed state may exhibit so-called frustration
because the packing - hexagonal for instance - cannot be perfect
(Earnshaw \& Harrison 1977; Pereira \& Williams 2000), but this
effect will be neglected here.

Let us then consider a configuration of a DNA molecule enclosed
within a capsid of volume $V_{0}$ and defined by the radius vector
$\vec{r}(s)$ as a function of the contour distance $s$ from one
end. The bending energy $U_{c}$ in a Hookean approximation is
written in terms of the local radius of curvature $R_{c}(s)$ of
the DNA curve at $s$ (Yamakawa 1971)
\begin{equation}
U_{c}=\frac{1}{2}Pk_{B}T\int_{0}^{L}\mbox{d}s\hspace{4pt}\frac{1}{R_{c}^{2}(s)}
\end{equation}
Here, the bending force constant equals $Pk_{B}T$ where $P$ is the
DNA persistence length, $k_{B}$ is Boltzmann's constant and $T$ is
the temperature. We wish to introduce the normalized density of
DNA segments at position $\vec{r}$ expressed in terms of a delta
function
\begin{equation}
\rho_{c}(\vec{r})=\frac{N}{L}\int_{0}^{L}\mbox{d}s\hspace{4pt}
\delta(\vec{r}-\vec{r}(s))
\end{equation}
\[
\int\mbox{d}\vec{r}\hspace{4pt}\rho_{c}(\vec{r})=N
\]
The contour length $L$ of the DNA consists of $N$ "segments" of
length $A$, the distance between elementary charges viewed along
the DNA helical axis. We next insert the density into eq (1) and
average over all configurations (denoted by $\langle\hspace{10pt}
\rangle$) which is possible because $\vec{r}(s)$ is a unique
nonintersecting curve.
\begin{eqnarray}
\langle U_{c}\rangle=U & = &
\frac{1}{2}PAk_{B}T\int\mbox{d}\vec{r}\mbox{
}\left\langle\frac{\rho_{c}(\vec{r})}{R_{c}^{2}(\vec{r})}\right\rangle
\nonumber
\\
&&\\ \nonumber & \simeq &
\frac{1}{2}PAk_{B}T\int\mbox{d}\vec{r}\mbox{
}\frac{\rho(\vec{r})}{R^{2}(\vec{r})}
\end{eqnarray}
We have also introduced a close-packing approximation
$\langle\rho_{c}(\vec{r})/R_{c}^{2}(\vec{r})\rangle\simeq\langle
\rho_{c}\rangle\langle R_{c}^{-2}\rangle\equiv\rho(\vec{r})
R^{-2}(\vec{r})$ which is certainly accurate when the DNA
undulations away from the average are small. I note that eq (3)
pertains to both purely mechanical and statistical mechanical
types of theories. Even in the former, we still need to average
over all configurations, for the capsid has a complicated shape so
there must be a host of DNA conformations of virtually identical
energies. The bending energy given by eq (3) was used recently
(Odijk and Slok 2003) to set up a simple density functional theory
of packing and will be commented on below.

Next, we consider a more severe approximation by letting the
density be constant. It is conveniently expressed in terms of the
area $S$ of the unit cell of the packaged DNA ($\rho_{0}\equiv
1/SA$) which we often take to be hexagonal. The supposition of
uniform density allows us to deduce expressions of some
generality. The bending energy is rewritten as
\begin{equation}
U=\frac{Pk_{B}T}{2S}\int_{V_{i}}^{V_{0}}\mbox{d}\vec{r}\hspace{4pt}R^{-2}(\vec{r})
\end{equation}
where the infinitesimal volume $\mbox{d}\vec{r}\equiv S\mbox{d}s$.
The volumes $V_{i}$ and $V_{0}$ are the inner and outer volumes,
respectively, enclosing the packed DNA (see fig 1). We assume
there is only one hole though it is stressed that general shapes
of the capsid and hole are allowed. Now it can be argued by
scaling and other arguments (Odijk 1986; Selinger \& Bruinsma
1991; Kamien et al 1992) that the twist and splay contributions to
the free energy of a tightly packed liquid crystal are small
compared with the bending energy. Thus we interpret $R^{-2}$ as an
energy density of pure bending which depends on the director
$\vec{n}(\vec{r})$ but not on $S$, the size of the unit cell.
Moreover, because there should be no hairpins at close packing,
the DNA mesophase is splayless at constant density (de Gennes
1977).

There is also a free energy of the DNA interacting with itself. We
omit minor curvature and surface contributions (Ubbink \& Odijk
1995) and express this energy as an extensive form proportional to
the contour length $L$
\begin{equation}
F_{int}=Lf(S)
\end{equation}
This pertains to a hypothetically straightened DNA lattice.
Accordingly, we now want to minimize the total free energy
\begin{equation}
F_{tot}=U+F_{int}
\end{equation}
subject to a volume constraint
\begin{equation}
V_{i}=V_{0}-LS
\end{equation}
We minimize $F_{tot}$ with respect to the director $\vec{n}$
(implicit in $R^{-2}$) and to the cell size $S$, keeping the outer
volume $V_{0}$ fixed but also the chemical potentials of the small
ions of the buffer in the reservoir containing the phage.
\begin{equation}
\left.\frac{\delta F_{tot}}{\delta\vec{n}}\right|_{S,V_{i}(S)}=0
\end{equation}
\[
\left.\frac{\partial F_{tot}}{\partial S}\right|_{\vec{n}}=0
\]
These expressions yield
\begin{equation}
Lf'(S)-\frac{U}{S}+\frac{P}{2S}\int_{A_{i}}\mbox{d}\vec{r}_{i}
\hspace{4pt}R_{m}^{-2}(\vec{r}_{i})\left|\frac{\partial\vec{l}(\vec{r}_{i})}
{\partial S}\right|=0
\end{equation}
in terms of an optimized inverse square radius of curvature
$R_{m}^{-2}$ and an integral over the surface of the inner hole
(see fig 2). A variation in $S$ induces a change in the vector
$\vec{l}$ at point $\vec{r}_{i}$, pointing normal to the inner
surface in the direction of the hole. The variation of the inner
volume can be similarly written with the help of eq (7)
\begin{equation}
\frac{\partial V_{i}}{\partial S}=-\int_{A_{i}}\mbox{d}\vec{r}_{i}
\hspace{4pt}\left|\frac{\partial\vec{l}(\vec{r}_{i})} {\partial
S}\right|=-L
\end{equation}
Hence, it is expedient to define a measure of the hole size given
by
\begin{equation}
E_{i}^{-2}\equiv\frac{\int_{A_{i}}\mbox{d}\vec{r}_{i}
\hspace{4pt}R_{m}^{-2}(\vec{r}_{i})\left|\frac{\partial\vec{l}(\vec{r}_{i})}
{\partial S}\right|}{\int_{A_{i}}\mbox{d}\vec{r}_{i}
\hspace{4pt}\left|\frac{\partial\vec{l}(\vec{r}_{i})} {\partial
S}\right|}
\end{equation}
Noting that $-f'(S)$ is simply the osmotic pressure $\pi_{os}$, we
finally attain the nanothermodynamic relation
\begin{equation}
\pi_{os}=-\frac{U}{SL}+\frac{Pk_{B}T}{2SE_{i}^{2}}
\end{equation}
The usual thermodynamics does not hold: if the hole is small, we
have $V_{0}={\mathcal O}(SL)$ and $U={\mathcal
O}(PV_{0}^{1/3}k_{B}T/S)$ and so the osmotic pressure is not a
purely intensive quantity.

Next, the average pressure ${\mathcal P}$ on the capsid wall
exerted by the DNA is computed by considering a virtual change in
$V_{0}$. Concomitantly, the director configuration
$\vec{n}(\vec{r})$, the inner volume $V_{i}(V_{0},S)$ and cell
size $S$ also change in such a way that the free energy remains a
minimum (eq 8) and the constraint (eq 7) is obeyed. The resulting
analysis is similar to that given above, leading to
\begin{equation}
{\mathcal P}=-\frac{\partial F_{tot}}{\partial V_{0}}= \left.
-\frac{\partial U}{\partial V_{0}}\right|_{S}=
\frac{Pk_{B}T}{2SE_{i}^{2}}-\frac{Pk_{B}T}{2SE_{0}^{2}}=
\frac{U}{SL}+\pi_{os}-\frac{Pk_{B}T}{2SE_{0}^{2}}
\end{equation}
The typical size $E_{0}$ of the capsid is defined via its surface
similarly to eq (11). I note that this pressure is not at all
identical to the osmotic pressure as one might have naively
surmised.

It is also interesting to study a quasistatic force ${\mathcal F}$
(in the absence of dissipative losses; Kindt et al 2001) upon
increasing the length of DNA in the capsid. We require local
equilibrium (eqs (14)) as $\vec{n}$, $S$ and $V_{i}(S,L)$ adjust
to the changing degree of packing subject to eq (13). It is
straightforward to show that the "pressure" ${\mathcal F}/S$ of
insertion is given by
\begin{equation}
\frac{\left|{\mathcal F}\right|}{S}=\frac{\partial
F_{tot}}{S\partial L}=\frac{f}{S}+\pi_{os}+\frac{U}{SL}
\end{equation}
Again, it would have been difficult to anticipate this expression.
We have derived eqs (12-14) for a phage of general shape at the
expense of introducing the capsid and hole sizes as unconventional
averages.

\section*{3. Hexagonal DNA lattice at high concentrations}

We have argued that we merely need the free energy per unit length
$f(S)$ of a straightened DNA lattice in order to evaluate
pertinent quantities of the DNA enclosed in a capsid. In this
section we assume such a lattice is perfectly hexagonal without
defects. Actually, the DNA chains also undulate about a reference
configuration but this effect is temporarily disregarded. The
lattice model for rodlike polyelectrolytes is, of course, well
known (Oosawa 1971), but the usual cell model which disposes of
the hexagonal symmetry, breaks down at high degrees of packing
when the polyion cylinders are close. A new approximation is
introduced here, though some general features of the Donnan
equilibrium need to be discussed first.

For now, the lattice is free from simple electrolyte and consists
of parallel cylindrical polyions together with attendant
counterions dissolved in the water in the intervening void. In
some model or approximation one introduces, there is generally an
interface or line of symmetry where the negative electrostatic
potential is a maximum and is set equal to zero. The electric
field vanishes so the osmotic pressure $\pi_{os}$ and chemical
potential of the ions $\mu_{i}$ reduce to (Israelachvili 1985)
\begin{eqnarray}
\pi_{os} & = & \bar{\rho}_{i}k_{B}T \nonumber\\
\mu_{i} & = & \mu_{ref}+k_{B}T\ln \bar{\rho}_{i}
\end{eqnarray}
where $\bar{\rho}_{i}$ is the concentration of counterions at the
interface. We restrict ourselves to monovalent ions bearing an
elementary charge $q$ which are ideal save for their Coulombic
interaction mediated by a supposedly uniform permittivity
$\epsilon$.

Next, the lattice is brought into thermodynamic contact with a
large reservoir containing monovalent salt of concentration
$c_{s}$. The activity coefficients of the small ions are set equal
to unity so the resulting Donnan equilibrium, due to the equality
of the chemical potentials of the small ions in the respective
phases, yields
\begin{equation}
\rho_{-}(\bar{\rho}_{i}+\rho_{+})=c_{s}^{2}
\end{equation}
The polyelectrolyte suspension contains positive and negative ions
arising from the simple salt, whose densities must be set equal in
view of electroneutrality $(\rho_{+}=\rho_{-})$. Therefore, upon
eliminating $\rho_{+}$ and $\rho_{-}$, the osmotic pressure is
simply (Oosawa 1971)
\begin{eqnarray}
\pi_{os} & = & (\bar{\rho}_{i}+\rho_{+}+\rho_{-}-2c_{s})k_{B}T
\nonumber\\
& = & \bar{\rho}_{i}k_{B}Tg(w)\\
g(y) & \equiv & (1+y^{2})^{1/2}-y \nonumber \\
w & \equiv & 2c_{s}/\bar{\rho}_{i} \nonumber
\end{eqnarray}
In the so-called cell model (developed by Fuoss, Katchalsky,
Oosawa and others, see Oosawa 1971), the counterion concentration
$\bar{\rho}_{i}$ at the interface in the salt-free case is given
by
\begin{equation}
\bar{\rho}_{i}\simeq\frac{A\rho_{0}}{2Q}
\end{equation}
This is valid at low and intermediate concentrations. For DNA, we
have a segment length $A=0.17$ nm and the Bjerrum length
$Q=q^{2}/\epsilon k_{B}T=0.71$ nm at room temperature implying
that a fraction of counterions appears to be associated with the
DNA rods. This is often termed counterion condensation (Manning
1969).

We now address the Poisson-Boltzmann equation in the lattice at
high concentrations (in the salt-free case). The ion distribution
$\rho(\vec{r})$ is connected to the (negative) electric potential
$\varphi$ via a Boltzmann distribution
\begin{equation}
\rho_{i}=q\bar{\rho}_{i}\exp(-q\varphi/k_{B}T)
\end{equation}
which is inserted in the Poisson equation
\begin{equation}
\Delta\varphi=-\frac{4\pi\rho_{i}}{\epsilon}
\end{equation}
This procedure is not entirely rigorous although it is quite
accurate for monovalent ions (the difficulty is that the two
densities in eqs (19) and (20) are very similar though not
identical; see Fixman 1979).

Let us consider the case when the surfaces of the polyion
cylinders are separated by only a short distance $h$ which is much
smaller than the cylinder radius $a$ (see fig 3). Hansen et al
(2001) then still insist on using a cell model in which the effect
of the rods surrounding a test rod is mimicked by a cylindrical
boundary. This seems a rather severe approximation. An alternative
is to view the hole enclosed within the triangle in fig 3 as a
separate entity which is feasible provided the electrostatic
screening is sufficiently high. Solving the Poisson-Boltzmann
equation in this geometry is still difficult, so we replace the
hole by three sections involving thin layers of electrolyte, and
one effective hollow cylinder whose radius $b$ is appropriately
chosen. I consider the latter first. The boundary condition for
the electric field at its surface is
\begin{equation}
\left.\frac{\mbox{d}\varphi}{\mbox{d}r}\right|_{b}=\frac{4\pi\sigma_{q}}{\epsilon}
\end{equation}
with $r$ the radial coordinate. The negative charge density
$\sigma_{q}=-q\sigma$ with $\sigma$ the number of charges per unit
area to be chosen below. We have to solve eq (19) and (20) so it
is convenient to introduce the dimensionless variables $\Psi\equiv
q\varphi/k_{B}T$ and $\bar{r}\equiv r/\lambda$ where the screening
length $\lambda$ is analogous to the Debye length
\begin{equation}
\lambda^{-2}\equiv 4\pi Q\bar{\rho}_{i}
\end{equation}
The counterion density $\bar{\rho}_{i}$ now pertains to the
centreline of the cylinder. There, the potential is set equal to
zero and the electric field is also zero so that eqs (15) hold. It
is then quadrature to solve the Poisson-Boltzmann equation:
\begin{equation}
\Psi=2\ln\left(1-\frac{1}{8}\bar{r}^{2}\right)
\end{equation}
The concentration $\bar{\rho}_{i}$ is determined from eqs (21) and
(22).
\begin{eqnarray}
\left(\frac{b}{\lambda}\right)^{2} & = &
\frac{8\pi\Lambda}{1+\pi\Lambda}\nonumber\\
&&\\ \Lambda & \equiv & Q\sigma b \nonumber
\end{eqnarray}
Hence, the osmotic pressure is conveniently expressed as
\begin{eqnarray}
\pi_{os} & = &
\bar{\rho}_{i}k_{B}T \nonumber \\
&&\\
& = & \frac{2\sigma
k_{B}T}{b}\left(\frac{1}{1+\pi\Lambda}\right)\nonumber
\end{eqnarray}
At small $\Lambda$, this reduces to a condition of
electroneutrality as it should, since no counterions are condensed
in this limit.

To progress, we need to set the variables $\sigma$ and $b$. The
size of the thin boundary layers discussed below is so small that
they may be neglected in the following argumentation. The
triangular region in fig 3 bears half an electron charge per
length $A$ along the major axis. The surface density of the
effective cylinder is thus $\sigma=1/4\pi bA$ implying that
$\Lambda\simeq 2$ is independent of the interaxial spacing $H$
($Q/A=4.2$ for DNA). The radius $b$ is chosen such that the
volumes of enclosed liquid are the same in the respective
compartments.
\begin{equation}
\pi b^{2}=\frac{1}{4}\sqrt{3}H^{2}-\frac{1}{2}\pi a^{2}
\end{equation}
We rewrite eq (25) by introducing the volume fraction
$v=\rho_{0}/\rho_{DNA}$ of DNA (the segment density of pure DNA
$=\rho_{DNA}$; the area of a hexagonal unit cell is
$S=\sqrt{3}H^{2}/2$).
\begin{eqnarray}
\pi_{os} & = & \bar{\rho}_{i}k_{B}T \nonumber\\
& = & \frac{\rho_{0}k_{B}T}{2(1-v)}
\end{eqnarray}
The density $\rho_{0}$ is identical to the uniform DNA density
discussed in the previous section.

Obviously, our choice of cylinder parameters is not at all
rigorous so the numerical coefficient in eq (27) is tentative. The
scaling of parameters in eq (25) is convincing, however, because
the correct limit given by eq (18) is attained at low densities
($H\gg a$ i.e. $\Lambda\gg 1$). Furthermore, the screening length
$\lambda$ is indeed small. For instance, at a separation $h$ very
close to zero, we have $b=0.23 a$ from eq (26) and since
$\bar{\rho}_{i}\simeq 1/4\pi b^{2}A$ we get $\lambda\simeq 0.47 a$
from eq (24). Hence, the respective triangular sections (fig 3) in
the entire lattice basically act as independent units as far as
the electrostatics is concerned.

The thin boundary layers are treated in a Derjaguin approximation
taking advantage of the relatively large radius ($a\gg \lambda$).
Of course, discrete charge effects play a role now, so our
analysis is not quantitatively accurate. For two flat plates of
charge density $\sigma=1/2\pi aA$ equal to that of DNA, we have
\begin{equation}
\pi_{os}=\frac{2\sigma k_{B}T}{h}
\end{equation}
The scaled density $\Lambda_{p}\equiv \sigma Qh$, the analogue of
$\Lambda$ in eq (24), is much smaller than unity so the
counterions are essentially uncondensed. The strategy of Derjaguin
is to superpose normal forces exerted by opposing infinitesimal
platelets on the two nearby charged objects (Israelachvili 1985).
We therefore focus on two particular cylinders in fig 3 and
introduce the plane defined by their centre axes. Two interacting
platelets are located at a height $x$ from this plane and the
distance $Z$ between them is $Z=h+2z$ with
$z=a-(a^{2}-x^{2})^{1/2}$. Hence, for relatively large $a$, we
have $Z\simeq h+(x^{2}/a)$. The normal force between the two
platelets per unit area is from eq (28)
\begin{equation}
{\mathcal F}_{p}(Z)=\frac{\mbox{2}\sigma k_{B}T}{Z}
\end{equation}
and the total force between the two cylinders of length $l$ is
hence given by
\begin{eqnarray}
{\mathcal F}(h) & \simeq & 2l\int_{Z=h}^{\infty}\mbox{d}
x\hspace{6pt}{\mathcal F}_{p}(Z) \nonumber\\
&&\nonumber\\ & = & \frac{4\sigma l
k_{B}T}{h}\int_{0}^{\infty}\mbox{d}x
\left(1+\frac{x^{2}}{ha}\right)^{-1}\nonumber\\
&&\nonumber\\
& = & 2\pi\sigma a^{1/2}h^{-1/2}l k_{B}T
\end{eqnarray}
This can be integrated to give $F_{2}$, the free energy of
interaction between the two cylinders. If there are $n$ rods in
the actual hexagonal lattice, the total free energy is
$F_{int}=3nF_{2}$ for there are six cylinders surrounding a test
rod though we have to avoid doublecounting. By definition the
osmotic pressure is
\begin{eqnarray}
\pi_{os,2} & = & -\frac{1}{3^{1/2}nLH}\frac{\partial
F_{int}}{\partial
H} \nonumber\\
&&\nonumber\\
& = & \frac{3^{1/2}k_{B}T}{AHa^{1/2}h^{1/2}}\nonumber\\
&&\nonumber\\
& = & \frac{3H\rho_{0}k_{B}T}{2a^{1/2}h^{1/2}}
\end{eqnarray}
upon using the density for DNA, $\sigma=1/2\pi aA$ and eq (30).

At small separations the density $\bar{\rho}_{i}$ in the middle of
the hollow cylinder (eq (27)) is the maximum density within the
triangle in fig. 3. There is another $\bar{\rho}_{i,2}$ defined
midway between the surfaces of two nearby cylinders. In accordance
with eq (28), this is given by $\bar{\rho}_{i,2}\simeq 1/\pi ahA$
so the screening length $\lambda=Aha/4Q$ from eq (22) is quite
small, significantly smaller, in fact, than the transverse scale
$a^{1/2}h^{1/2}$ of the "surface of interaction" evident in the
integrand of eq (30). This justifies the Derjaguin approximation
but also the splitting up of the original triangular region into a
hollow cylinder and three thin boundary layers. We now place the
lattice in a reservoir containing monovalent electrolyte of
concentration $c_{s}$. A Donnan equilibrium is established as
described by eqs (16) and (17) with $\bar{\rho}_{i}$ given by eq
(27).

\section*{4. Friction within condensed DNA}

As the phage connector forces the DNA genome into a capsid, the
DNA coil is bound to twist in view of its helical backbone. Since
the other end probably also becomes constrained at some stage, we
expect a twist-to-writhe transition to occur in dealing with the
DNA dynamics. At first, the DNA density is statistically isotropic
more or less, although the density decreases towards the central
region owing to the curvature stress (Odijk \& Slok 2003). As the
concentration is enhanced, the excluded-volume effect, at some
point, causes the DNA globule to become (probably) nematic in the
outer confines of the capsid while the inner region remains
isotropic (I disregard the chirality of the DNA interaction for
simplicity). As more DNA is piled up into the phage, the entire
globule could become nematic in a spoollike configuration. At the
next stage, a hexagonal (or hexatic) phase appears, again, in the
outer region in coexistence with a nematic inner phase.
Ultimately, at high degrees of packing, the entire structure
becomes hexagonal apart from faults caused by frustration. This
scenario is tentative of course. An a priori theory of globular
organization is formidable as noted earlier. One problem is how
the DNA loops are forced into ordered phases though we expect
rippling transitions to occur (Odijk 1996).

I here present two highly simplified notions of the DNA dynamics.
At low degrees of congestion, the usual reptation dynamics (de
Gennes 1979) may be thought to apply, where a test chain slithers
through a tubelike environment. Gabashvili and Grosberg (1992)
already considered reptation during the ejection of DNA from
phages. The friction the DNA experiences as it is transported
through the connector, was also accounted for. Here, the reptation
of DNA is discussed in the context of cyclization in order to
gauge the accuracy of such a picture. At high degrees of
congestion, the osmotic stress builds up because of the strong
curvature stress. Simple reptation no longer seems likely then so
I introduce a model based on sliding friction.

\vspace{24pt}

\noindent{\it Cyclization within globules}

The DNA is compacted into a globule by external forces (e.g. by
adding inert polymer or multivalent ions, by confinement within a
capsid in the initial stages). As argued by Ubbink and Odijk
(1995, 1996), it is often possible to view the globular shape as
being determined by balancing bending against surface forces while
keeping the volume $V_{t}$ fixed. For an ideal toroid defined by
the two radii $R_{r}$ and $R_{t}$ shown in fig 4, the surface area
is $A_{t}=4\pi^{2}R_{r}R_{t}$ and the volume
$V_{t}=SL=2\pi^{2}R_{r}^{2}R_{t}$ is also given in terms of the
(constant) area $S$ of the unit cell. The two ends would never
come into contact if the shape of the toroid were to remain the
same, for the free energy of change in volume is prohibitive. We
thus investigate fluctuations in the globular shape. I desregard
free energy contributions arising from crossovers, knots and other
defects (Park et al 1998; Arsuaga et al 2002; Hud \& Downing 2001;
Nelson 2002).

The curvature stress is low if $R_{t}\gtrsim 2 R_{r}$ (Ubbink \&
Odijk 1995; Ubbink \& Odijk 1996). Under these conditions the
bending energy (eq (1)) is approximated by
\begin{equation}
U\simeq\frac{PL}{2R_{t}^{2}}
\end{equation}
There is a surface tension $\sigma$ scaled by $k_{B}T$ whatever
the mechanism of compaction. The surface free energy is then
\begin{equation}
F_{S}=\sigma A_{t}k_{B}T
\end{equation}
Since $V_{t}$ is constrained, we eliminate one of the radii so as
to express the total free energy in units of $k_{B}T$  as
\begin{equation}
F_{t}=\frac{1}{2}PLR_{t}^{-2}+2^{3/2}\pi\sigma
R_{t}^{1/2}V_{t}^{1/2}
\end{equation}
Minimization of $F_{t}$ with respect to $R_{t}$ yields an optimum
radius of the toroid
\begin{equation}
R_{t,m}^{5/2}=\frac{PL}{2^{1/2}\pi\sigma V_{t}^{1/2}}
\end{equation}
Therefore, the fluctuations in globular shape are governed by a
Gaussian  distribution
\begin{eqnarray}
G & \propto & \exp-\frac{1}{2}F_{t}''\Delta R_{t}^{2} \\
&&\nonumber \\ \left\langle\Delta R_{t}^{2}\right\rangle & \equiv
& \left\langle\left(R_{t}-
R_{t,m}\right)^{2}\right\rangle=\left(F_{t}''\right)^{-1}
\nonumber \\
&&\nonumber \\
F_{t}'' & = & \frac{5}{2}PLR_{t,m}^{-4} \nonumber
\end{eqnarray}
This is derived by a straightforward Taylor expansion of eq (34)
within the Boltzmann factor $\exp(-F_{t})$.

We next need to assess how the winding number $n$ changes in
response to alterations in shape. This is simply $nS=\pi
R_{r}^{2}$ so a fluctuation in $n$ is given by $\delta n=2\pi
R_{r,m}\delta R_{r}/S$. The volume constraint implies
$2R_{r,m}R_{t,m}\delta R_{c}+R_{r,m}^{2}\delta R_{t}=0$. If the
two ends are to meet, we require $\delta n\geq 1$, leading to an
inequality for the second moment (via eq (36)). We finally attain
the condition $L>10\pi^{2}P$ which, remarkably, is independent of
the surface tension. Accordingly, by this mechanism, cyclization
can only occur if the DNA is quite long.

A reptation time for cyclization is now readily written down by
dimensional reasoning. It must be proportional to the friction
exerted by the fluid (water presumably, with a viscosity
$\eta_{0}$, since the globule is not very compact), i.e.
proportional to the contour length $L$. For cyclization, one of
the DNA ends has to diffuse a distance, typically about $\pi
R_{t}$ at least, with respect to the other end.
\begin{equation}
\tau_{r}\simeq\frac{(\pi R_{t})^{2}L\eta_{0}}{k_{B}T}
\end{equation}
We do not address the dynamics of the two ends as they approach
each other within a region of size $\sim R_{r}$.

\vspace{24pt}

\noindent{\it Sliding friction in highly compacted states}

Let us first consider a classic example of Coulomb friction
(Persson 2000). A block of some material rests on a flat surface
(fig 5). Owing to gravity, for instance, the block exerts a
pressure ${\mathcal F}_{S}/\Omega$ on a section of the surface of
area $\Omega$. If we now exert a force ${\mathcal F}_{t}$ on the
block, tangential to the surface, the blocks fails to move at
first. It is only at a certain ${\mathcal F}_{t}$ given by
\begin{equation}
\frac{{\mathcal F}_{t}}{{\mathcal F}_{s}}= \frac{{\mathcal
F}_{t}/\Omega}{{\mathcal F}_{s}/\Omega}=\omega
\end{equation}
that the block is set in motion. The friction constant $\omega$ is
well defined, at least within certain bounds on the velocity thus
induced.

In a highly congested DNA globule compressed by bending stresses,
we may also discern a pressure $\pi_{os}$ acting on a surface area
$2\sqrt{3}LH$ where $2\sqrt{3}H$ is the length of the perimeter of
the unit cell assumed to be hexagonal. Hence, the minimum force
needed to set the DNA in motion should be
\begin{eqnarray}
{\mathcal F}_{t} & = & 2\sqrt{3}\omega HL\pi_{os}\nonumber\\
&&\\
& \simeq & \frac{2\omega LPk_{B}T}{HE_{i}^{2}}\nonumber
\end{eqnarray}
At very high degrees of packing, the first term in eq (12) may be
neglected. The minimum force increases fast upon further packaging
of DNA in a phage (increasing $L$, decreasing $H$ and $E_{i}$). It
has been assumed that the DNA slides homogeneously through its own
tubelike environment.

\section*{5. Discussion}

\noindent{\it Ordered phases in the bulk and in phages}

It is important to summarize several results for
liquid-crystalline DNA in the bulk that are relevant to DNA
packaged within phages. The literature on DNA liquid crystals in
vitro is vast and has been reviewed elsewhere (Livolant \&
Leforestier 1996). Durand et al (1992) performed careful X-ray
diffraction experiments on slowly evaporating suspensions of 50 nm
long DNA rods. The sequence of phase transitions was argued to be:
isotropic - cholesteric - 2d hexagonal - 3d hexagonal -
orthorhombic. The transitions are first order except for the one
between the two hexagonal phases. In the 2d hexagonal phase, the
rods are free to move longitudinally but this freedom is gradually
frozen out as the DNA concentration increases, so the 2d-3d
transition is continuous. The solution only becomes orthorhombic
at a very high concentration of 670 g/l (the interaxial spacing
$H=2.37$ nm in the coexisting 3d hexagonal phase). The helical
repeat actually decreases quite dramatically from 3.5 nm to 3.0 nm
within the sequence of positionally ordered phases, whereas the B
conformation remains surprisingly intact with a constant rise $=2A
=0.336$ nm. The buffer contained 0.25 M electrolyte (either
ammonium acetate or sodium chloride) so these data are pertinent
to the packing of DNA under physiological conditions.

The ionic-strength dependence of the cholesteric-hexagonal
transition for DNA of length almost 50 nm was monitored by
Kassapidou et al (1998), the hexagonal spacings at coexistence
ranging from 3.7 nm at 0.10 M NaCl to 3.0 nm at l.65 M. The
orientational fluctuations measured by neutron scattering were
close to 10 degrees. The first-order transition is visible, in
fact, in a test tube with a clear meniscus discernible between the
two phases and the hexagonal phase exhibits the characteristic
fanlike texture under a polarizing microscope (J.R.C. van der
Maarel, private communication, 1998).

On the other hand, Strey et al (2000) do not seem to witness the
X-ray diffraction characteristic of hexagonal order, for
concentrated Calf-Thymus DNA in 0.5 M NaCl, that has been
wet-spun. One potential drawback is that the long DNA is quite
polydisperse which could prevent a phase transition applicable to
very long monodisperse DNA. Anyway, the azimuthal scattering has
sixfold symmetry so that the phase is probably hexatic. If so, the
decay in the orientational order should be algebraic (Nelson 2002)
with an exponent depending on the conditions (like the ionic
strength). This ought to be gleaned from the scattering curves
(Lyuksyutov et al 1992).

Spool models have been adopted as models of DNA packaged in
bacteriophages (Harrison 1973). X-ray diffraction patterns from
phages have often been interpreted in terms of hexagonal packing
of the DNA (North \& Rich 1961; Earnshaw \& Harrison 1977; Stroud
et al 1981; Ront\'{o} et al 1988; Booy et al 1991; Cerritelli et
al 1997; Bhella et al 2000). Cerritelli et al (1997) have adduced
quite convincing evidence for the validity of the spool form for
well-aligned bacteriophage T7. Earnshaw and Harrison (1977)
pointed out that the inner region of the DNA spool should be more
disordered. A theory of the nonuniform density of the DNA inside
phages was set up recently (Odijk \& Slok 2003) in which a
cholesteric-hexagonal transition could be exhibited even under
conditions of rather tight packing (note the spacings quoted
above). Physicists have analyzed the X-ray scattering by nematic
liquid crystals of poly-$\gamma$-benzyl glutamate in terms of
density fluctuations (Ao et al 1991). These are coupled  to
director fluctuations by so-called continuity laws (de Gennes
1977, Selinger \& Bruinsma 1991, Kamien et al 1992). The intensity
spectrum as a function of the scattering vector has a
characteristic bowtie shape. There is a typical length scale
associated with these fluctuations which is determined by a
trade-off of bending versus osmotic compressibility (de Gennes
1977). This turns out to be similar in size to $E_{i}$ the radius
of the inner hole of the spool inside phages (bulk hexagonal DNA
here being compared to phage DNA). Thus, it would be of some
interest to reinvestigate the scattering by the inner region of
the DNA spool from this point of view (Odijk \& Slok 2003).

The osmotic pressure of DNA solutions has also been readdressed
recently (Strey et al 1999, Raspaud et al 2000). At very high
concentrations where the phase is hexagonal or hexatic (but I
shall assume it to be hexagonal for definiteness), there are
significant deviations from the Oosawa theory (eqs (17) and (18)).
The approximate theory in terms of a hollow cylinder and thin
boundary layers goes some way towards explaining the high values
of $\pi_{os}$ (Strey et al 1999). For instance, at a DNA
concentration of 1.2 monoM, eq (27) predicts $2.4\times 10^{6}$
N/m$^{2}$ in good agreement with experiment (eq (31) does not
apply since the separation is $h={\mathcal O}(a)$ and not $h\ll
a$). At a concentration of 1.8 monoM, eqs (27) and (31) yield
osmotic pressures $5.2\times 10^{6}$ N/m$^{2}$ and $13\times
10^{6}$ N/m$^{2}$ respectively, compared with the experimental
$13\times 10^{6}$ N/m$^{2}$ (Strey et al 1999).

Till now, we have neglected configurational entropy. The free
energy of a wormlike chain trapped in a tube of diameter less than
the persistence length $P$ was formulated long ago (Odijk 1983)
and has been discussed many times since. For DNA confined within a
hexagonal lattice, we may write on the basis of scaling arguments
(Odijk 1983, Selinger \& Bruinsma 1991)
\begin{equation}
f_{en}(S)=\frac{c_{1}}{h^{2/3}P^{1/3}}
\end{equation}
\begin{equation}
\pi_{os}=\frac{2c_{1}}{3^{3/2}h^{5/3}HP^{1/3}}
\end{equation}
Unfortunately, the coefficient $c_{1}$ is not at all known
accurately for a hexagonal phase. Setting the coefficient
$c_{1}=1$ and using eqs (7), (12) and (14) lead to a
semiquantitative explanation of the fairly sharp rise in the
loading force as an elastic worm becomes densely packed in a
cavity (see the simulations of  Kindt et al (2001) and of
Marenduzzo and Micheletti (2003). The latter authors also find
that there is a section of chain aligned within the cylindrical
hole of the spool. Arsuaga et al (2002) put forward evidence for
an inner region misaligned with the well-ordered outer spool under
conditions of intermediate packaging but the effects of realistic
electrostatics on these computations need to be awaited.) For
comparison, the osmotic pressure of the entropic interaction is
$5.3\times 10^{5}$ N/m$^{2}$ at 1.8 monoM, an order of magnitude
smaller than the pressures quoted above ($h=0.51$ nm, $P=50$ nm).
Nevertheless, the nonlinear coupling between electrostatics and
undulatory entropy probably needs looking into (see the work of De
Vries (1997) on undulating charged surfaces in the salt-free
Poisson-Boltzmann approximation).

\vspace{24pt}

\noindent{\it Loading forces and inner radii}

At high degrees of packing, the terms containing $U$ in eqs (12),
(13) and (14) are negligible so the relevant thermodynamic
parameters reduce to ${\mathcal P}\simeq\pi_{os}\simeq
Pk_{B}T/2SE^{2}$ and $|{\mathcal F}|=f+S\pi_{os}$. The hole inside
a spoollike region may have caps belonging to the wall of the
capsid, a fact disregarded in section 2, but this effect gives
only corrections of ${\mathcal O}
\left(E_{i}^{2}/V_{0}^{2/3}\right)$. We thus regain eq (10) of
Odijk (1998) valid for a specific model, which itself is in good
agreement with the DNA spacings measured inside bacteriophage T7
by Cerritelli at al (1997), Odijk (1998) employed the simple
Oosawa model (eqs (17) and (18)). In recent papers, Tzlil et al
(2002) and Purohit et al (2003) have used a different form:
$\pi_{os}=\pi_{0}[\exp(H_{0}-H)/\lambda_{H}-1]$ with
$\pi_{0}=0.12$ $k_{B}T/$nm$^{3}$, decay length $\lambda_{H}=0.14$
nm and $H_{0}=2.8$ nm stemming from the osmotic stress
measurements of  Rau \& Parsegian (1992) on DNA greatly perturbed
by trivalent cobalt hexammine ions. Why this interaction,
attractive in part, should generally prevail inside phages is not
clear. Anyway, the spacings predicted  by eqs (7) and (12) upon
inserting this form, are virtually identical to those in table 1
of Tzlil et at (2002). The inner radius $E_{i}$ is now 2.81 nm
when the entire genome is packaged.

The supposition of uniform density has been released in a recent
analysis based on eq. (3) (Odijk \& Slok 2003). In a density
functional approach, the gradient terms are small because they
exert themselves on a scale $\lambda$ or $H$ only. The curvature
stress is balanced against the chemical potential of inhomogeneous
bulk DNA. The hole size is given by an expression similar to eq
(12) but the tightly wound region surrounding the hole has an
inhomogeneous density varying on a scale $E_{i}$ (Odijk \& Slok
2003). Such a profile seems to be in qualitative agreement with
experiments of Olson et al (2001) on the T4 phage.

The forces of encapsidation have been monitored in single-molecule
experiments by Smith  et al (2001). I shall try to interpret this
force semiempirically. It seems a good Ansatz to set the osmotic
pressure $\pi_{os}/k_{B}T=\bar{\rho}_{i}=c_{2}\rho_{0}=c_{2}/AS$
within a certain interval of the DNA concentration in the absence
of salt (see eqs (27) and (31) here, and fig 2 of Hansen et al
(2001)). We then need the quantity $f(S)$ at a certain ionic
strength of the buffer. There is a Donnan equilibrium given by eq
(17) and the free energy has been computed by Odijk \& Slok
(2003). We therefore obtain for the force $|{\mathcal F}
|=f(S)+S\pi_{os}=(c_{2}k_{B}T/A)[1-\ln w]$ with
$w=2c_{s}/\bar{\rho}_{i}=2ASc_{s}/c_{2}\ll 1$. Smith et al (2001)
measured a maximum loading force of 50 pN at a buffer
concentration $c_{s}$ of about 0.1 M. The inner volume of the
phi29 capsid can be approximated by an ellipsoid of major axis
equal to 25.5 nm and minor axis equal to 19.5 nm (Tao et al 1998;
P.J. Jardine, private communication, 2000). Hence we have
$V_{0}=40600$ nm$^{3}$ and so the unit cell has dimensions
$S\simeq 6.2$ nm$^{2}$ and $H\simeq 2.7$ nm for a genome length
$L_{max}$ if we disregard the very small inner hole of size
$E_{i}=2.4$ nm, computed from eq (12) upon a first iteration. We
finally calculate $c_{2}=0.73$ and the osmotic pressure $\pi_{os}$
is about 29 atm. This underestimates the theoretical and
experimental bulk pressures a bit but the effect of magnesium ions
in the force set-up has not been accounted for.

However, there are difficulties in the interpretation of the
force-length curve below the limit of full packing. If the inner
hole remains fairly small, we may express the derivative as
\begin{equation}
\frac{\partial |{\mathcal F}|}{\partial L}\simeq
-\frac{S^{2}}{L}\frac{\partial \pi_{os}}{\partial S}
\end{equation}
using eqs (7) and (14). Accordingly, $\partial |{\mathcal F}
|/\partial L=c_{2}k_{B}T/AL=18$ pN per $L_{max}$ if
$\pi_{os}=c_{1}\rho_{0}$. This is not unreasonable if $L\simeq
L_{max}$ (see Smith et al 2001) but fails utterly when
$0.4<L/L_{max}<0.9$, for $\partial |{\mathcal F} |/\partial L$
increases upon compaction in that case. It appears that we must
have the osmotic pressure increasing as $\rho_{0}^{2}$ at least.
What could be the cause of such a high effective power, granting
that the measured force does not reflect dissipative losses?
First, the crossover concentration demarcating the validity of the
Oosawa model vis-\`{a}-vis a model with complete hexagonal detail
is about 0.8 monoM judging from fig 2 of Hansen et al (2001)
($L=0.5 L_{max}$ at this concentration). Thus, counterions are
gradually decondensing within the anomalous range of contour
lengths. In addition, the Donnan effect is also slowly diminishing
and the effect of the thin boundary layers is also becoming
appreciable. An increase of the pressure stronger than linear is
not surprising though this issue merits further quantitative
study. (Parenthetically, a problem possibly related to this could
be the curvature in the rate-length curves determined by P.J.
Jardine, opposite to the one measured by Smith et al (2001) (P.J.
Jardine, private communications, 2000, 2001). He packaged DNA in
phi29 phages in vitro by adding ATP to the solution. The reaction
was effectively blocked with the help of gamma-S-ATP and the
length of DNA injected was quantified as a function of time using
a restriction enzyme.)

The inner radii computed above at a high degree of packing are
very small. An important problem that now arises is whether such
tiny radii are viable, both physically and biologically. We are
here concerned with the tight curvature of long sections of
solenoidal DNA independent of the base pair sequence and in the
absence of bound proteins. Crick and Klug (1975) already quoted
values of 3 to 5 nm for radii of curvature that could pose
potential problems with regard to smooth bending. Sussman \&
Trifonov (1978) estimated a critical radius of  kinking of about
2.5 nm by equating the elastic bending energy to the maximum
enthalpy of stacking the base pairs assuming one kink per helical
repeat. A more detailed treatment based on Van der Waals contacts
led to a minimum acceptable radius of about 5 nm. More recently, a
51 bp microcircle with a radius of curvature of $2.8$ nm was
devised by exploiting current computer modelling techniques (Tung
\& Soumpasis 1995). 42 bp microcircles have actually been
constructed in the laboratory (Wolters \& Wittig 1989, 1990) but
the base pair sequence is from a promotor. On the whole, a viable
minimum radius would thus seem to be 3 nm or only very slightly
smaller (A. Travers, private communication, 2003). The inner radii
$E_{i}$ within phages that we have computed above are dangerously
small. It is possible that the energy of curvature becomes
anharmonic at small radii which would increase estimates of
$E_{i}$ somewhat. Any kinking would, of course, render invalid the
models outlined here.

\hspace{24pt}

\noindent{\it DNA self-friction}

Jary and Sikorav (1999) investigated the monomolecular collapse of
lambda DNA caused by the trivalent polyamine spermidine and the
attendant rate of cyclization of the DNA. Fortunately, the
collapse is very fast so it is possible to monitor the cyclization
in the toroidal globular state quite accurately. The measured
rates were between about 1 to 10 $s^{-1}$. This compares
favourably with the reptation time from eq (37) (the viscosity of
water $\eta_{0}=1$ cP at room temperature, the contour length
$L=17$ $\mu$m and $R_{t}\simeq 60$ nm (Widom \& Baldwin 1983)). We
conclude that the DNA apparently experiences friction from the
"water" alone, despite the fact that the typical spacing in such a
globule is small ($H\simeq 2.8$ nm).

In a tightly packed phage, the dynamics of the DNA could be
markedly different from that in globules, because of the much
higher bending stress. In evaluating the frictional forces between
two surfaces separated by a very thin layer of intervening fluid,
one introduces a dimensionless group, $\eta_{0}u/\pi_{os}$ in this
case, where $\vec{u}$ is the velocity tangential to the surface
(see fig 5; Persson 2000). In DNA encapsidation within a phage,
this velocity is so small (Smith et al 2001), that this quantity
is minute so we ought te have boundary lubrication rather than
hydrodynamic lubrication. The friction coefficient $\omega$ would
be about $10^{-1}$ (Persson 2000). In principle, there should then
be a tendency for the water to be squeezed  out between two
adjacent DNA strands. Because there are only a few water molecules
between the strands, layering transitions would be possible
(Zilberman et al 2001) depending on the degree of commensurability
of the two DNA backbones (M\"{u}ser et al 2001). However, eq (39)
would now predict a minimum force more than 100 times larger than
the 50 pN force observed in the phi29 phage. A possible resolution
of this paradox is the very recent assessment of sliding friction
between surfaces with intervening aqueous nanolayers (Raviv et al
2001, Raviv \& Klein 2002). Surprisingly, the water still acts
almost like a hydrodynamic bulk fluid, possibly because the
hydrogen bonds are suppressed under conditions of sliding
nanofriction. The friction coefficient could now be as low as
$10^{-3}$ so the 50 pN force (Smith et al 2001) may be just enough
to overcome the minimum force. It is also remarked that the free
energy of compacting DNA into the phi29 phage is about 1 $k_{B}T$
per bp (Smith et al 2001) to be compared with the enthalpy of
expulsion, 0.80 $k_{B}T$ per bp, measured by calorimetry for
bacteriophage T7 (Raman et al 2001; Serwer 2003), though the
packing conditions in these two phages are not identical.
Obviously, the problem of dissipative losses, if any, merits
further attention.

\section*{6. Concluding Remarks}

Clearly, our understanding of DNA packing in bacteriophages is
incomplete. Although we may rationalize the high osmotic pressures
and loading forces as being caused by the small separation between
DNA helices, the tight inner radii of the spools are too close to
kinking for comfort. At intermediate degrees of packing, we lack a
precise theory of the electrostatic interaction in terms of the
fraction of counterions condensed on the DNA. The loading forces
are very sensitive to detail in the interaction in this regime. An
open problem is the friction at a high degree of compaction and
its nature - is it hydrodynamic lubrication? Lastly, magnesium
ions are often present in the buffers used. Their impact on the
Donnan equlibrium needs to be investigated.

\section*{Acknowledgment}

I am grateful to Jean-Louis Sikorav, Edouard Yeramian and Paul
Jardine for discussions, and the latter for extended
correspondence on phages and packaging. Professor Andrew Travers
is thanked for very helpful thoughts on the minimum radius of
curvature of DNA. The insights of Flodder Slok proved most
welcome.

\newpage

\section*{references}

\begin{description}
\item[]Ao , X., Wen, X. \& Meyer, R.B. 1991 X-ray scattering from
polymer nematic liquid crystals. \textit{Physica A} \textbf{176},
63-71.

\item[]Arsuaga, J., V\'{a}zquez, M., Trigueros, S., De Witt
Sumners \& Roca, J. 2002 Knotting probability of DNA molecules
confined in restricted volumes: DNA knotting in phage capsids.
\textit{Proc. Natl. Acad. Sci. U.S.A.} \textbf{99}, 5373-5377.

\item[]Arsuaga, J., Tan R.K.Z., V\'{a}zquez, M., De Witt Sumners
\& Harvey, S.C. 2002 Investigation of viral DNA packaging using
molecular mechanics models. \textit{Biophys. Chem.}
\textbf{101-102}, 475-484.

\item[]Bhella, D., Rixon, F.J. \& Dargan, D.J. 2000 Cryomicroscopy
of human cytomegalovirus virions reveals more densely packed
genomic DNA than in herpes simplex virus type 1. \textit{J. Mol.
Biol.} \textbf{295}, 155-161.

\item[]Booy, F.P., Newcomb, W.W., Trus, B.L., Brown, J.C., Baker,
T.S. \& Steven, A.C. 1991 Liquid crystalline, phage-like packing
of encapsidated DNA in herpes simplex virus. \textit{Cell}
\textbf{64}, 1007-1015.

\item[]Cerritelli, M.E., Cheng, N., Rosenberg, A.H., McPherson,
C.E., Booy, F.P. \& Steven, A.C. 1997 Encapsidated conformation of
bacteriophage T7 DNA. \textit{Cell} \textbf{91}, 271-280.

\item[]Crick, F.H.C. \& Klug, A. 1975 Kinky helix. \textit{Nature}
\textbf{255}, 530-533.

\item[]de Gennes, P.G. 1977 Polymeric liquid crystals: Frank
elasticity and light scattering. \textit{Mol. Cryst. Liq. Cryst.}
\textbf{34}, 177-182.

\item[]de Gennes, P.G. 1979 \textit{Scaling concepts in polymer
physics.} Cornell University Press, Ithaca, N.Y.

\item[]de Vries, R. 1997 Thermal undulations in salt-free charged
lamellar phases: theory versus experiment. \textit{Phys. Rev. E}
\textbf{56}, 1879-1886.

\item[]Durand, D., Doucet, J. \& Livolant, F. 1992 A study of the
structure of highly concentrated phases of DNA by X-ray
diffraction. \textit{J. Phys. II France 2} \textbf{9}, 1769-1783.

\item[]Earnshaw, W.C. \& Harrison, S.C. 1977 DNA arrangement in
isometric phage heads. \textit{Nature} \textbf{268}, 598-602.

\item[]Fixman, M. 1979 The Poisson-Boltzmann equation and its
application to polyelectrolytes. \textit{J. Chem. Phys.}
\textbf{70}, 4995-5005.

\item[]Gabashvili, I.S., Grosberg, A.Yu., Kuznetsov, D.V. \&
Mrevlishvili, G.M. 1991 Theoretical model of DNA packing in the
phage head. \textit{Biophysics} \textbf{36}, 782-789

\item[]Gabashvili, I.S. \& Grosberg, A.Yu. 1992 Dynamics of double
stranded DNA reptation from bacteriophage. \textit{J. Biomol.
Struct. Dyn.} \textbf{9}, 911-920.

\item[]Grimes, S., Anderson, D.L., Baker, T.S. \& Rossmann, M.G.
2000 Structure of the bacteriophage phi29 DNA packaging motor.
\textit{Nature} \textbf{408}, 745-750.

\item[]Grimes, S., Jardine, P.J. \& Anderson, D. 2002
Bacteriophage phi29 DNA packaging. \textit{Adv. Vir. Res.}
\textbf{58}, 255-294.

\item[]Hansen, P.L., Podgornik, R. \& Parsegian, V.A. 2001 Osmotic
properties of DNA: critical evaluation of counterion condensation
theory. \textit{Phys. Rev. E} \textbf{64}, 021907 (1-4).

\item[]Harrison, S.C. 1983 Packaging of DNA into bacteriophage
heads: A model. \textit{J. Mol. Biol.} \textbf{171}, 577-580

\item[]Hud, N.V. \& Downing, K.H. 2001 Cryoelectron microscopy of
lambda phage DNA condensates in vitreous ice: The fine structure
of DNA toroids. \textit{Proc. Natl. Acad. Sci. USA} \textbf{98},
14925-14931.

\item[]Israelachvili, J.N. 1985 \textit{Intermolecular and surface
forces}, Academic, London.

\item[]Jardine, P.J. \& Anderson, D. 2003 DNA packaging. In
\textit{The Bacteriophages}, Ed. R. Calendar, Oxford University
Press, in press.

\item[]Jary, D.\& Sikorav, J.-L. 1999 Cyclization of  globular
DNA. Implications for DNA-DNA interactions in vivo.
\textit{Biochemistry} \textbf{38}, 3223-3227.

\item[]Kamien, R.D., Le Doussal, P. \& Nelson, D.R. 1992 Theory of
directed polymers. \textit{Phys. Rev. A} \textbf{45}, 8727-8750.

\item[]Kassapidou, K. \& Van der Maarel, J.R.C. 1998 Melting of
columnar hexagonal DNA liquid crystals. \textit{Eur. Phys. J. B.}
\textbf{3}, 471-476.

\item[]Kindt, J., Tzlil, A., Ben-Shaul, A. \& Gelbart, W.M. 2001
DNA packaging and ejection forces in bacteriophage. \textit{Proc.
Natl. Acad. Sci. USA} \textbf{98}, 13671-13674.

\item[]Livolant, F. \& Leforestier, A. 1996 Condensed phases of
DNA: structure and phase transitions. \textit{Prog. Polym. Sci.}
\textbf{21}, 1115-1164.

\item[]Lyuksyutov I., Naumovets, A.G. \& Pokrovsky, V. 1992
\textit{Two-dimensional crystals Academic}, San Diego, CA.

\item[]Manning, G.S. 1969 Limiting laws and counterion
condensation in polyelectrolyte solutions. I. Colligative
properties. \textit{J. Chem. Phys.} \textbf{51}, 924-933

\item[]Marenduzzo, D. \& Micheletti, C. 2003 Thermodynamics of DNA
packaging inside a viral capsid: the role of DNA intrinsic
thickness, preprint.

\item[]M\"{u}ser, M.H., Wenning, L. \& Robbins, M.O. 2001 Simple
microscopic theory of Amontons's laws for static friction.
\textit{Phys. Rev. Lett.} \textbf{86}, 1295-1298.

\item[]Nelson, D.R. 2002 \textit{Defects and geometry in condensed
matter physics}. Cambridge University Press, U.K.

\item[]North, A.C.T. \& Rich, A. 1961 X-ray diffraction studies of
bacterial viruses. \textit{Nature} \textbf{191}, 1242-1247.

\item[]Odijk, T. 1983 On the statistics and dynamics of confined
or entangled stiff polymers. \textit{Macromolecules} \textbf{16},
1340-1344.

\item[]Odijk, T. 1986 Elastic constants of nematic solutions of
rodlike and semiflexible polymers. \textit{Liquid Crystals}
\textbf{1}, 553-559.

\item[]Odijk, T. 1996 DNA in a liquid-crystalline environment:
tight bends, rings, supercoils. \textit{J. Chem. Phys.}
\textbf{105}, 1270-1286.

\item[]Odijk, T. 1998 Hexagonally packed DNA within bacteriophage
T7 stabilized by curvature stress. \textit{Biophys.J.}
\textbf{75}, 1223-1227.

\item[]Odijk, T. \& Slok, F. 2003 Nonuniform Donnan equilibrium
within bacteriophages packed with DNA. \textit{J. Phys. Chem. B.}
in press.

\item[]Olson, N.H., Gingery, M., Eiserling, F.A. \& Baker, T.S.
2001 The structure of isometric capsids of bacteriophage T4.
\textit{Virology} \textbf{279}, 385-391.

\item[]Oosawa, F. 1971 \textit{Polyelectrolytes}, Marcel Dekker,
New York.

\item[]Park, S.Y., Harries, D. \& Gelbart, W.M. 1998 Topological
defects and the optimum size of DNA condensates.
\textit{Biophys.J.} \textbf{75}, 714-720.

\item[]Pereira, G.G. \& Williams, D.R.M. 2000 Crystalline toroidal
globules of DNA and other semiflexible polymers: jumps in radius
caused by hexagonal packing. \textit{Europhys. Lett.} \textbf{50},
559-564.

\item[]Persson, B.N.J. 2000 \textit{Sliding friction}. Springer,
Berlin, 2nd ed.

\item[]Purohit, P.K., Kondev, J. \& Phillips, R. 2003 Mechanics of
DNA packaging in viruses. \textit{Proc. Natl. Acad. Sci. USA}
\textbf{100}, 3173-3178.

\item[]Raman, C.S., Hayes, S.J., Nall, B.T. \& Serwer, P. 2001
Energy stored in the packaged DNA of bacteriophage T7.
\textit{Biophys. J.} \textbf{65}, A12.

\item[]Raspaud, E., Da Concei\c{c}ao, M. \& Livolant, F. 2000 Do
free DNA counterions control the osmotic pressure? \textit{Phys.
Rev. Lett.} \textbf{84}, 2533-2536.

\item[]Rau, D.C. \& Parsegian, V.A. 1992 Direct measurement of the
intermolecular forces between counterion-condensed DNA double
helices. \textit{Biophys. J.} \textbf{61}, 246-259.

\item[]Raviv, U., Laurat, P. \& Klein, J. 2001 Fluidity of water
confined to subnanometre films. \textit{Nature} \textbf{413},
51-54.

\item[]Raviv, U. \& Klein, J. 2002 Fluidity of bound hydration
layers. \textit{Science} \textbf{297}, 1540-1543.

\item[]Reimer, S.C. \& Bloomfield, V.A. 1978 Packaging of DNA in
bacteriophage heads: some considerations on energetics.
\textit{Biopolymers} \textbf{17}, 785-794.

\item[]Ront\'{o}, Gy., T\'{o}th, K., Feigin, L.A., Svergun, D.I.
\& Dembo, A.T. 1988 Symmetry and structure of bacteriophage T7.
\textit{Comput. Math. Applic.} \textbf{16}, 617-628.

\item[]Selinger, J.V. \& Bruinsma, R.F. 1991 Hexagonal and nematic
phases of chains. I. Correlation functions. \textit{Phys. Rev. A}
\textbf{43}, 2910-2921.

\item[]Serwer, P. 2003 Models of bacteriophage DNA packaging
motors. \textit{J. Struct. Biol.} \textbf{141}, 179-188.

\item[]Simpson, A.A., Tao, Y., Leiman, P.G., Badasso, M.O., He Y.,
Jardine, P.J., Olson, N.H., Morals, M.C., Grimes, S., Anderson,
D.L., Baker, T.S. \& Rossmann, M.G. 2000 Structure of the
bacteriophage phi29 DNA packaging motor. \textit{Nature}
\textbf{408}, 745-750.

\item[]Smith, D.E. Tans, S.J., Smith, S.B., Grimes, S., Anderson,
D.L. \& Bustamante, C. The bacteriophage phi29 portal motor can
package DNA against a large internal force. \textit{Nature}
\textbf{413}, 748-752.

\item[]Strey, H.H., Parsegian, V.A. \& Podgornik, R. 1999 Equation
of state for polymer liquid crystals: Theory and experiment.
\textit{Phys. Rev. E} \textbf{59}, 999-1008.

\item[]Strey, H.H., Wang, J., Podgornik, R., Rupprecht, A., Yu,
L., Parsegian, V.A. \& Sitora, E.B. 2000 Refusing to twist:
Demonstration of a line hexatic phase in DNA liquid crystals.
\textit{Phys. Rev. Lett.} \textbf{84}, 3105-3108.

\item[]Stroud, R.M., Serwer, P. \& Ross, M.J. 1981 Assembly of
bacteriophage T7. Dimensions of the bacteriophage and its capsids.
\textit{Biophys. J.} \textbf{36}, 743-757.

\item[]Sussman, J.L. \& Trifonov, E.N. 1978 Possibility of
nonkinked packing of DNA in chromatin. \textit{Proc. Natl. Acad.
Sci. USA} \textbf{75}, 103-107.

\item[]Tao, Y., Olson, N.H., Xu, W., Anderson, D.L., Rossmann,
M.G. \& Baker, T.S. 1998 Assembly of a tailed bacterial virus and
its genome release studied in three dimensions. \textit{Cell}
\textbf{95}, 431-437.

\item[]Tung, C.S. \& Soumpasis, D.M. 1995 The construction of DNA
helical duplexes along prescribed 3-D curves. \textit{J. Biomol.
Struct. Dyn.} \textbf{13}, 577-582.

\item[]Tzlil, S., Kindt, J.T. Gelbart, W.M. \& Ben-Shaul, A. 2003
Forces and pressures in DNA packaging and release from viral
capsids. \textit{Biophys. J.} \textbf{84}, 1616-1627.

\item[]Ubbink, J. \& Odijk, T. 1995 Polymer- and salt-induced
toroids of hexagonal DNA. \textit{Biophys. J.} \textbf{68}, 54-61.

\item[]Ubbink, J. \& Odijk, T. 1996 Deformation of toroidal DNA
condensates under surface stress. \textit{Europhys. Lett.}
\textbf{33}, 353-358.

\item[]Widom, J. \& Baldwin, R.L. 1983 Monomolecular condensation
of lambda-DNA induced by cobalt hexammine. \textit{Biopolymers}
\textbf{22}, 1595-1620.

\item[]Wolters, M. \& Wittig, B. 1989. Construction of a 42 base
pair double stranded DNA microcircle. \textit{Nucl. Acids Res.}
\textbf{17}, 5163-5172.

\item[]Wolters, M. \& Wittig, B. 1990. A strategy for the
construction of double-stranded DNA microcircles with
circumferences less than 50 bp. \textit{Anti-Cancer Drug Design}
\textbf{5}, 111-120.

\item[]Yamakawa, H. 1971 \textit{Modern theory of polymer
solutions.} Harper \& Row, New York.

\item[]Zilberman, S., Persson, B.N.J., Nitzan, A., Mugele, F. \&
Salmeron, M. 2001 Boundary lubrication: dynamics of squeeze-ont.
\textit{Phys. Rev. E} \textbf{63}, 055103 (R) (1-4).

\end{description}

\newpage

\begin{figure}[htbp]
  \begin{center}
  \leavevmode
  \includegraphics[width=4in]{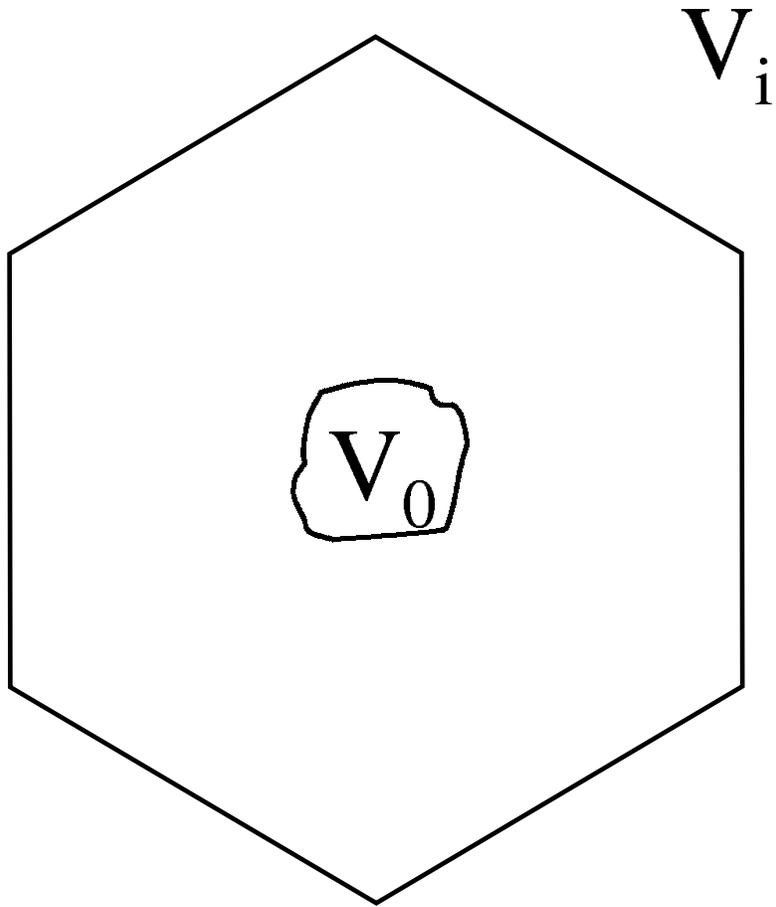}
  \end{center}
  \caption{A capsid packed with DNA  down to an inner
hole of volume $V_{0}$.}
\end{figure}
\newpage

\begin{figure}[htbp]
  \begin{center}
  \leavevmode  
  \includegraphics[width=5in]{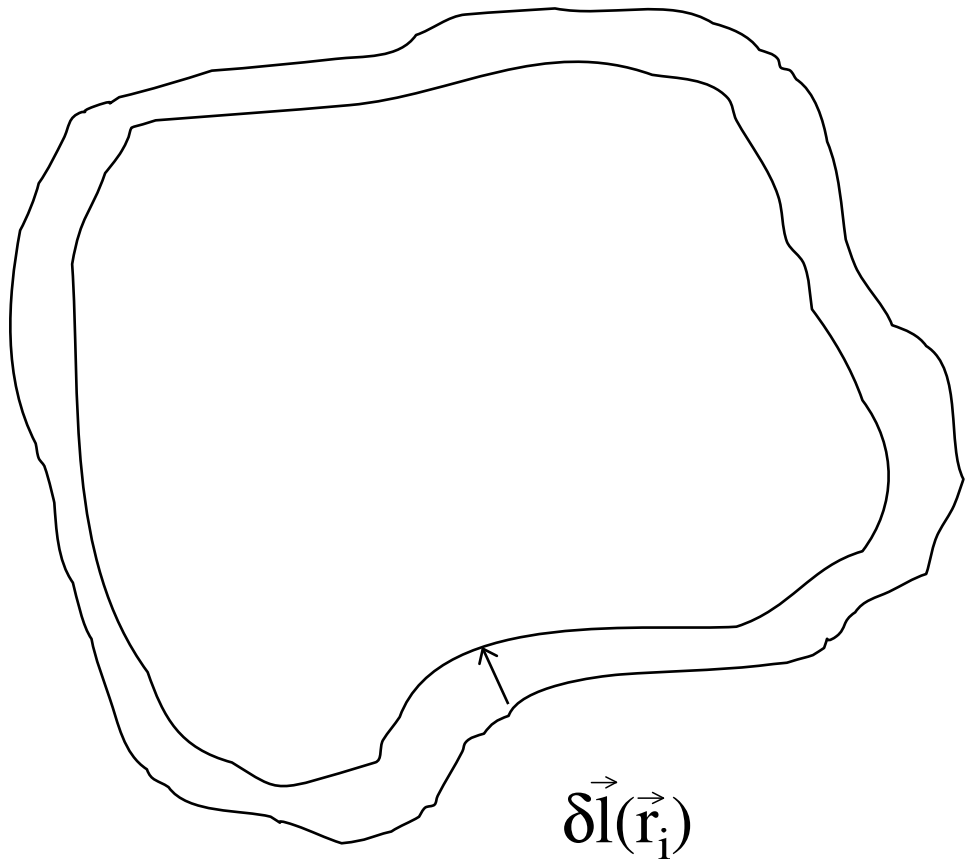}
  \end{center}
  \caption{The variational decrease in the size of the
hole within the DNA spool.}
\end{figure}
\newpage

\begin{figure}[htbp]
  \begin{center}
  \leavevmode
 \includegraphics[width=5in]{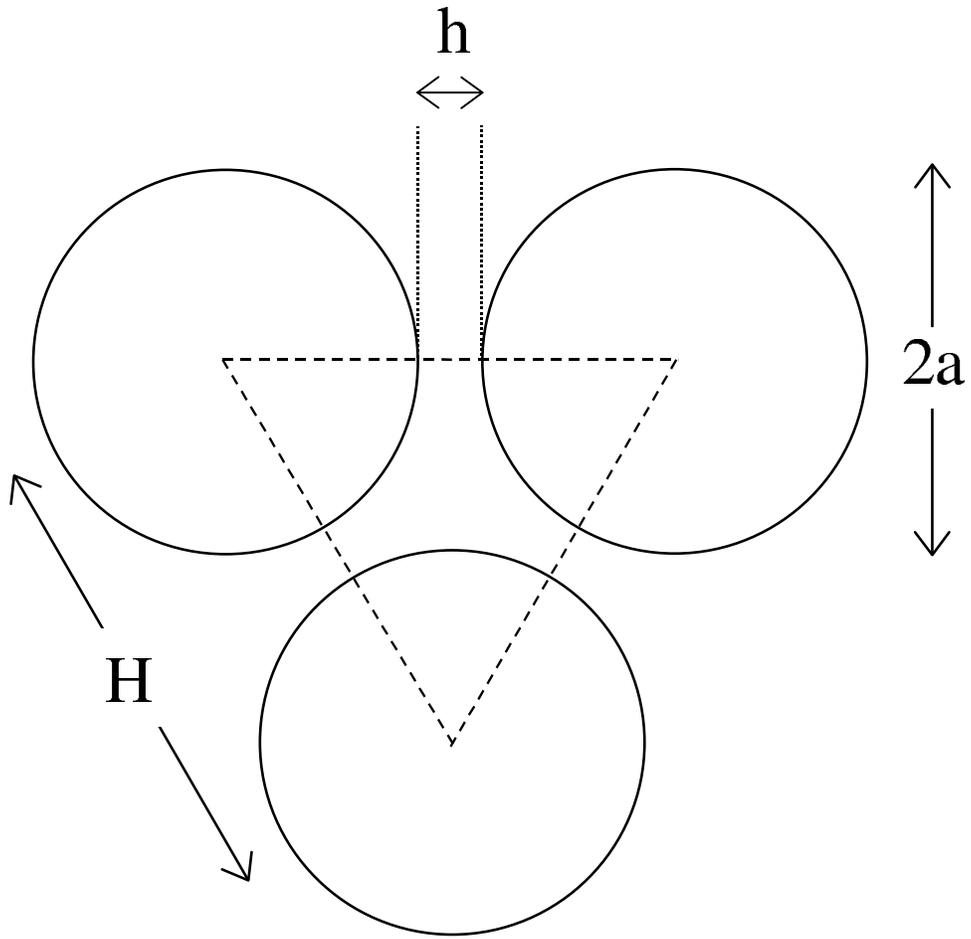}
  \end{center}
  \caption{Three DNA cylinders within the hexagonal
lattice.}
\end{figure}
\newpage

\begin{figure}[htbp]
  \begin{center}
  \leavevmode
 \includegraphics[width=5in]{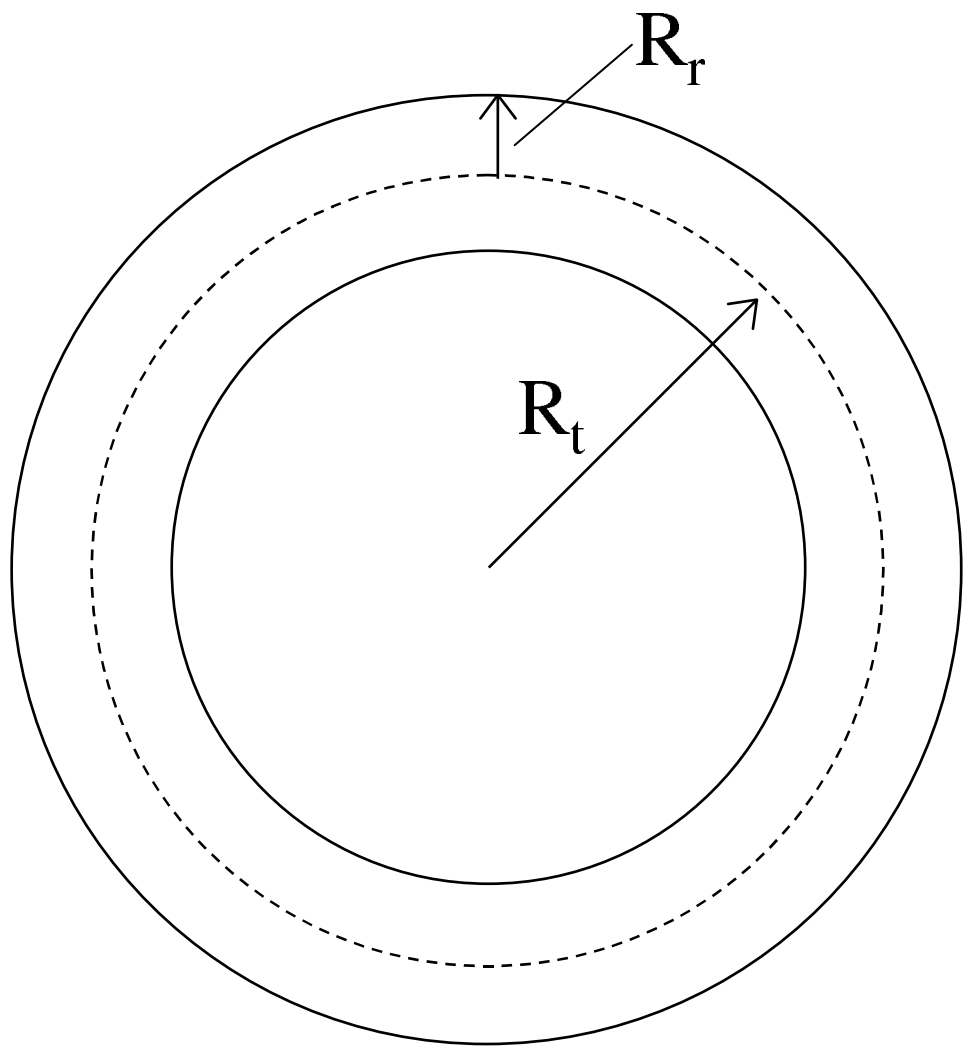}
  \end{center}
  \caption{Cross section of the toroidal DNA globule.}
\end{figure}
\newpage

\begin{figure}[htbp]
  \begin{center}
  \leavevmode
 \includegraphics[width=5in]{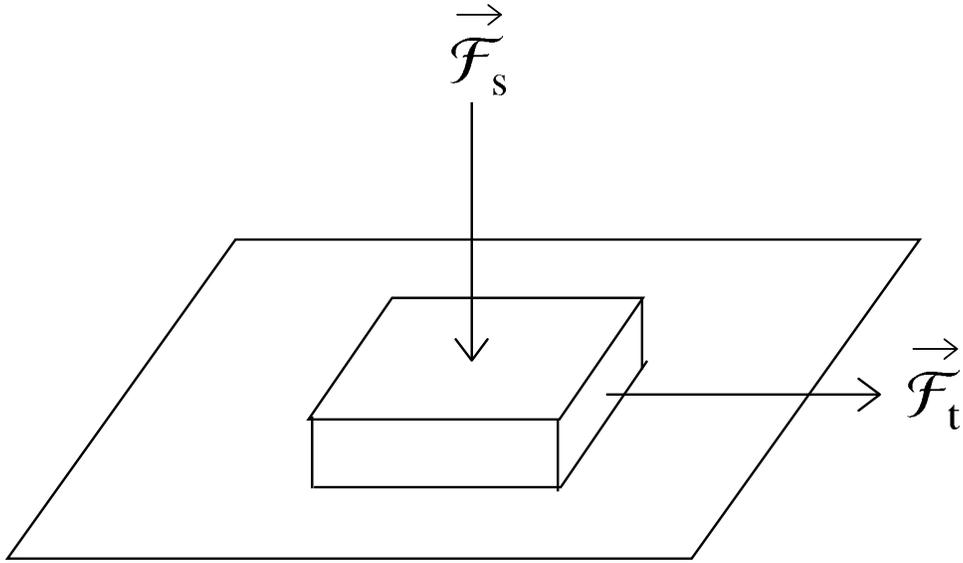}
  \end{center}
  \caption{A standard arrangement to illustrate Coulomb
friction.}
\end{figure}

\end{document}